\documentclass[english,12pt]{article}
          \usepackage{babel}
          \usepackage[T1]{fontenc}
          \usepackage[latin2]{inputenc}
          \usepackage{pslatex}
\begin{document}

\title{\bf Curious Variables Experiment (CURVE). \\
RZ LMi - the most active SU UMa star.}
\author{A. ~O~l~e~c~h$^1$, ~M. ~W~i~{\'s}~n~i~e~w~s~k~i$^1$, 
~K. ~Z~{\l}~o~c~z~e~w~s~k~i$^1$, ~L.M. ~C~o~o~k$^2$, \\
~K. ~M~u~l~a~r~c~z~y~k$^3$, ~and~ ~P. ~K~\c{e}~d~z~i~e~r~s~k~i$^3$}
\date{$^1$ Nicolaus Copernicus Astronomical Center,
Polish Academy of Sciences,
ul.~Bartycka~18, 00-716~Warszawa, Poland\\
{\tt e-mail: (olech,mwisniew,kzlocz)@camk.edu.pl}\\
~\\
$^2$ Center for Backyard Astrophysics (Concord), \\ 1730 Helix Court, Concord,
CA 94518, USA \\ {\tt e-mail: lcoo@yahoo.com}\\
~\\
$^3$ Warsaw University Observatory, Al. Ujazdowskie 4, 00-476 Warszawa,
Poland\\ {\tt e-mail: (kmularcz,pkedzier)@astrouw.edu.pl}}
\maketitle

\begin{abstract} 

We report extensive photometry of the frequently outbursting dwarf nova
RZ Leo Minoris. During two seasons of observations we detected 12
superoutbursts and 7 normal outbursts. The $V$ magnitude of the star
varied in range from 16.5 to 13.9 mag. The superoutbursts occur quite
regularly flashing every 19.07(4) days and lasting slightly over 10
days. The average interval between two successive normal outbursts is
4.027(3) days. The mean superhump period observed during the
superoutbursts is $P_{\rm sh}=0.059396(4)$ days ($85.530\pm0.006$ min).
The period of the superhumps was constant except for one superoutburst when
it increased with a rate of $\dot P/P_{\rm sh} = 7.6(1.9)\cdot
10^{-5}$. Our observations indicate that RZ LMi goes into long intervals
of showing permanent superhumps which are observed both in
superoutbursts and quiescence. This may indicate that decoupling of
thermal and tidal instabilities play important role in ER UMa systems.
No periodic light variations which can be connected with orbital period
of the binary were seen, thus the mass ratio and evolutionary status of
RZ LMi are still unknown.

\end{abstract}

\section{Introduction}

Dwarf novae are believed to be unmagnetized close binary systems
containing white dwarf primary and low mass main sequence secondary. The
secondary fills its Roche lobe and looses the material through the inner
Lagrangian point. This matter forms an accretion disc around the white
dwarf.

One of the most intriguing classes of dwarf novae are SU UMa stars which
have short orbital periods (less than 2.5 hours) and show two types of
outbursts: normal outbursts and superoutbursts. Superoutbursts are
typically about one magnitude brighter than normal outbursts, occur
about ten times less frequently and display characteristic tooth-shape
light modulations i.e. so called superhumps (see Warner 1995 for
review).

The behavior of SU UMa stars in now quite well understood within the
frame of the thermal-tidal instability model (see Osaki 1996 for
review). Superhumps occur at a period slightly longer than the orbital
period of the binary system. They are most probably the result of
accretion disk precession caused by gravitational perturbations from the
secondary. These perturbations are most effective when disk particles
moving in eccentric orbits enter the 3:1 resonance. Then the superhump
period is simply the beat period between orbital and precession rate
periods. Although in the last decades significant progress has been
made in explaining the behaviour of dwarf novae light curves, some
physical processes ongoing in these systems are still not fully
understood (see for example Smak 2000, Schreiber and Lasota 2007).

In the beginning of 90ties of XX century, SU UMa stars were believed to
be quite uniform group of variables with common properties. These
objects went into superoutburst every year or so and between two
successive superoutbursts showed $\sim$10 ordinary outbursts. However,
there were some exceptions like WZ Sge, which show infrequent and large
amplitude superoutburst followed by the period of quiescence with no
single eruption lasting even 30 years.

In 1995 astronomical community was alerted about the presence of stars
characterized by complete opposite behavior. First, Misslet and Shafter
(1995) reported observations of PG 0943+521 (later called ER UMa), which
allowed to detect superhumps with period of 0.0656 days and include this
object into the SU UMa group of variables. The most intriguing feature
of the long term light curve of ER UMa was very short interval between
two successive superoutbursts (so called supercycle) reaching only 44
days. This value was about three times shorter than shortest previously
known supercycles. This work was quickly followed by paper of Robertson
et al. (1995), who confirmed all findings of Misslet and Shafter (1995)
and precisely determined the value of supercycle of ER UMa to be equal
to 42.95 days. Moreover, they found two more objects with similar
properies - V1159 Ori with supercycle of 44.5 days and RZ LMi with
supercycle as short as 18.87 days! In the same year Nogami et al. (1995)
published paper which confirmed extremely short supercycle of RZ LMi and
showing that it belongs to SU UMa variables exhibiting clear superhumps
with period of 0.05946 days.

One year later the number of these unusual variables increased to four
objects. Kato et al. (1996) reported the discovery that DI UMa has a
supercycle of 25 days and shows clear superhumps with period of 0.0555
days.

The fifth ER UMa-type variable - IX Dra - was discovered by Ishioka et
al. (2001). Their observations revealed a supercycle length of 53 days
and an interval between normal outbursts of 3-4 days. Olech et al. (2004)
determined precisely both superhump and orbital periods of the binary
and estimated the supercycle length to 54 days.

The basic properies of five known up-today members of ER UMa group are
summarized in Table 1.

\begin{table}[!h]
\caption{\sc Basic properties of ER UMa variables. $P_{\rm orb}$ and $P_{\rm sh}$ denote
orbital and superhump periods, $\epsilon$ is a period excess, $T_s$ and $T_n$ are supercycle
and cycle periods, $T_{\rm sup}$ is duration of the superoutburst, $A_{\rm sup}$ and $A_{\rm n}$
are amplitudes of superoutburst and normal outburst.}
\vspace{0.1cm}
\begin{center}
\begin{tabular}{|l|c|c|c|c|c|c|c|c|c|}
\hline
\hline
Star & $P_{\rm orb}$ & $P_{\rm sh}$ & $\epsilon$ & $T_s$ & $T_n$ &
$T_{\rm sup}$ & $A_{\rm sup}$ & $A_{\rm n}$ & Ref\\
  & [days] & [days] & [\%] & [days] & [days] & [days] & [mag] & [mag] & \\
\hline
\hline
RZ LMi    & ?        & 0.05946  &  ?   & 18.9      & 3.8 & 11 & 2.5 & 2.0 & (1,2)\\
DI UMa    & 0.054564 & 0.0555   & 1.72 & 25.0      & 5.0 & 12 & 2.9 & 2.1 & (3,4)\\
ER UMa    & 0.06366  & 0.065552 & 2.97 & 43.0      & 4.4 & 23 & 2.6 & 2.2 & (2,5,6)\\
V1159 Ori & 0.062178 & 0.064284 & 2.11 & 44.6-53.3 & 4.0 & 16 & 2.2 & 1.4 & (2,6,7,8)\\
IX Dra    & 0.06646  & 0.066968 & 0.76 & 54.0      & 3.1 & 16 & 2.2 & 1.7 & (9,10)\\  
\hline
\hline
\multicolumn{10}{l}{\small 1. Nogami et al. (1995), ~2. Robertson et al. (1995), 
~3. Kato et al. (1996), ~4. Thorstensen et al. (2002)}\\
\multicolumn{10}{l}{\small 5. Kato et al. (2003), ~6. Thorstensen et al. (1995),
~7. Kato (2001), ~8. Patterson et al. (1995)}\\
\multicolumn{10}{l}{\small 9. Ishioka et al. (2001), ~10. Olech et al. (2004)}\\
\end{tabular}
\end{center}
\end{table}
\bigskip

It is clear that ER UMa stars consist a group of variables with common
properties such as extremely short supercycles, small amplitudes of
eruptions and relatively long superoutbursts lasting even longer than
half of the supercycle. However, the period excess $\epsilon$ defined as
$P_{\rm sh}/P_{\rm orb}-1$, which is connected with mass ratio by the
following relation:

\begin{equation}
\epsilon\approx\frac{0.23q}{1+0.27q}
\end{equation}

\noindent suggests different evolutionary status of the particular
members of ER UMa stars. For example, ER UMa and V1159 Ori seem to have
normal secondaries and evolve towards the shorter orbital periods. On
the other hand, DI UMa and IX Dra are much more evolved objects with
sub-stellar secondaries (possibly degenerate brown dwarfs) and evolve
towards the longer orbital periods (Patterson 2001, Olech et al. 2004).
The question why DI UMa and IX Dra are so active, while WZ Sge stars
having similar period excesses have longest supercycles, is still open.

\section{RZ Leo Minoris}

The variability of RZ LMi was discovered by Lipovetskij and Stepanjan
(1981). Spectra obtained by Green et al. (1982) suggested that the star
is dwarf nova with broad hydrogen and helium emission lines and with
clear variability in $B$ filter in the range from 16.8 to 14.4 mag. Another
spectrum obtained by Szkody and Howell (1992) showed H$\beta$ and
H$\gamma$ absorption features and H$\alpha$ absorption with emission
core, which indicated the presence of an accretion disk at high
mass-transfer rate. RZ LMi was also included as a cataclysmic variable
candidate in Palomar-Green Survey (Green et al. 1986) and designated as
PG 0948+344.

Long term CCD photometry spanning over 2.5 years of almost continuous
observations was presented by Robertson et al. (1995). The light curve
in $V$ was characterized by long eruptions repeating quite regularly
every 18.87 days and short eruptions occurring every 3.8 days. The
brightness of the star varied from 17.0 to 14.2 mag.

Photometry made by Nogami et al. (1995) confirmed extremely short
supercycle of RZ LMi and allowed to precisely determine the superhump
period as equal to 0.05946(4) days.

The extreme properties of RZ LMi, its relatively high brightness, lack
of precise photometry in minimum light, unknown orbital period and
determination of the superhump period basing on the data from only one
superoutburst encouraged us to include this object into the list of
variables regularly monitored within the Curious Variables Experiment
(Olech et al. 2003, 2006).

~

\begin{table}[!ht]
{\tiny
\caption{\sc Journal of the CCD observations of RZ LMi.}
\begin{center}
\begin{tabular}{|l|c|c|r|r|}
\hline
\hline
Date   & No. of & Start & End & Length \\
       & frames & 2453000. + & 2453000. + & [hr]~ \\
\hline
2004 Jan 22/23 & 2 & 27.49164 & 27.49409 & 0.059\\
2004 Jan 25/26 & 52 & 30.39285 & 30.50032 & 2.579\\
2004 Jan 26/27 & 19 & 31.59037 & 30.66757 & 1.853\\
2004 Jan 29/30 & 89 & 34.35335 & 34.71332 & 4.192\\
2004 Jan 30/31 & 75 & 35.29448 & 35.52361 & 4.578\\
2004 Feb 01/02 & 12 & 37.25596 & 37.29419 & 0.918\\
2004 Feb 11/12 & 70 & 47.32035 & 47.62186 & 4.664\\
2004 Feb 12/13 & 33 & 48.27176 & 48.61116 & 8.146\\
2004 Feb 16/17 & 12 & 52.66542 & 52.69522 & 0.715\\
2004 Feb 19/20 & 198 & 55.26513 & 55.67128 & 9.748\\
2004 Feb 20/21 & 240 & 56.26647 & 56.69513 & 10.288\\
2004 Feb 21/22 & 144 & 57.25898 & 57.48494 & 5.423\\
2004 Feb 24/25 & 115 & 60.45026 & 60.68871 & 5.723\\
2004 Feb 25/26 & 29 & 61.57348 & 61.65343 & 1.919\\
2004 Feb 26/27 & 96 & 62.27642 & 62.50980 & 5.601\\
2004 Mar 10/11 & 77 & 75.40324 & 75.62008 & 5.204\\
2004 Mar 11/12 & 134 & 76.26170 & 76.60946 & 6.371\\
2004 Mar 12/13 & 139 & 77.26967 & 77.55541 & 6.858\\
2004 Mar 13/14 & 126 & 78.24668 & 78.54966 & 7.272\\
2004 Mar 17/18 & 54 & 82.26416 & 82.51271 & 3.205\\
2004 Mar 18/19 & 38 & 83.31747 & 83.42568 & 2.597\\
2004 Mar 19/20 & 38 & 84.29448 & 84.42922 & 3.234\\
2004 Mar 21/22 & 48 & 86.27288 & 86.39105 & 2.836\\
2004 Mar 22/23 & 53 & 87.26646 & 87.37461 & 2.596\\
2004 Mar 29/30 & 21 & 94.45288 & 94.49385 & 0.983\\
2004 Mar 30/31 & 46 & 95.31690 & 95.40032 & 2.002\\
2004 Apr 13/14 & 44 & 109.34651 & 109.50887 & 3.897\\
2004 Apr 14/15 & 108 & 110.28399 & 110.49866 & 5.152\\
2004 Apr 15/16 & 66 & 111.27972 & 111.40584 & 3.027\\
2004 Apr 16/17 & 45 & 112.33495 & 112.43971 & 2.514\\
2004 Apr 18/19 & 24 & 114.42736 & 114.47021 & 1.028\\
2004 Apr 19/20 & 69 & 115.28831 & 115.42555 & 3.294\\
2004 Apr 20/21 & 116 & 116.27826 & 116.44833 & 4.082\\
2004 Apr 21/22 & 72 & 117.28336 & 117.42132 & 3.311\\
2004 Apr 22/23 & 92 & 118.28635 & 118.45840 & 4.129\\
2004 Apr 23/24 & 16 & 119.28807 & 119.33030 & 1.014\\
2004 Apr 25/26 & 14 & 121.28508 & 121.30873 & 0.568\\
2004 May 03/04 & 54 & 129.33305 & 129.46404 & 3.144\\
2004 May 04/05 & 49 & 130.33826 & 130.43341 & 2.284\\
2004 May 05/06 & 67 & 131.32664 & 131.44949 & 2.949\\
2004 May 10/11 & 2 & 136.43389 & 136.43620 & 0.055\\
2004 May 11/12 & 41 & 137.33082 & 137.41001 & 1.901\\
2004 May 12/13 & 45 & 138.32363 & 138.42612 & 2.460\\
2004 May 14/15 & 37 & 140.33032 & 140.39889 & 1.646\\
2004 May 16/17 & 10 & 142.33325 & 142.38237 & 1.179\\
2004 May 17/18 & 25 & 143.33488 & 143.39415 & 1.422\\
2004 May 23/24 & 29 & 149.35765 & 149.41197 & 1.307\\
2004 May 24/25 & 2 & 150.38973 & 150.39133 & 0.038\\
2005 Jan 06/07 & 31 & 377.55144 & 377.63966 & 0.088\\
2005 Jan 10/11 & 45 & 381.62064 & 381.70004 & 0.079\\
2005 Jan 16/17 & 33 & 387.65040 & 387.72563 & 0.075\\
2005 Jan 31/01 & 44 & 402.47627 & 402.56624 & 0.090\\
2005 Feb 07/08 & 115 & 409.37052 & 409.66598 & 0.295\\
2005 Feb 08/09 & 60 & 410.40827 & 410.56327 & 0.155\\
2005 Feb 09/10 & 115 & 411.34197 & 411.68835 & 0.346\\
2005 Feb 10/11 & 28 & 412.23383 & 412.35866 & 0.125\\
2005 Feb 11/12 & 46 & 413.51103 & 413.68581 & 0.175\\
2005 Feb 28/01 & 12 & 430.36068 & 430.39882 & 0.038\\
2005 Mar 03/04 & 97 & 433.30174 & 433.51610 & 0.214\\
2005 Mar 19/20 & 37 & 449.34525 & 449.43875 & 0.094\\
2005 Mar 29/30 & 98 & 459.30230 & 459.54215 & 0.240\\
2005 Mar 30/31 & 36 & 460.38366 & 460.53344 & 0.150\\
2005 Mar 31/01 & 85 & 461.29092 & 461.53520 & 0.244\\
2005 Apr 01/02 & 106 & 462.27025 & 462.57886 & 0.309\\
2005 Apr 02/03 & 111 & 463.31098 & 463.53601 & 0.225\\
2005 Apr 03/04 & 54 & 464.27473 & 464.41983 & 0.145\\
2005 Apr 04/05 & 68 & 465.26230 & 465.46418 & 0.222\\
2005 Apr 04/05 & 35 & 465.79699 & 465.88925 & 0.092\\
2005 Apr 05/06 & 45 & 466.27872 & 466.36908 & 0.090\\
2005 Apr 05/06 & 56 & 466.71948 & 466.86403 & 0.144\\
2005 Apr 06/07 & 22 & 467.33496 & 467.39442 & 0.059\\
2005 Apr 09/10 & 66 & 470.70411 & 470.86280 & 0.159\\
2005 Apr 12/13 & 153 & 473.66064 & 473.85721 & 0.197\\
2005 Apr 13/14 & 143 & 474.27636 & 474.49374 & 0.217\\
2005 Apr 14/15 & 177 & 475.65247 & 475.85657 & 0.204\\
2005 Apr 15/16 & 186 & 476.64940 & 476.88060 & 0.231\\
2005 Apr 17/18 & 150 & 478.66522 & 478.83682 & 0.172\\
2005 Apr 19/20 & 73 & 480.67021 & 480.83888 & 0.169\\
2005 Apr 20/21 & 61 & 481.65846 & 481.80204 & 0.144\\
2005 Apr 28/29 & 7 & 489.45682 & 489.47262 & 0.016\\
2005 May 07/08 & 34 & 498.32302 & 498.41290 & 0.090\\
2005 May 10/11 & 40 & 501.31475 & 501.41301 & 0.098\\
2005 May 11/12 & 3 & 502.36084 & 502.36564 & 0.005\\
2005 May 20/21 & 36 & 511.34924 & 511.41113 & 0.062\\
2005 May 21/22 & 44 & 512.34993 & 512.42105 & 0.071\\
2005 May 28/29 & 13 & 519.38338 & 519.39926 & 0.016\\
\hline
TOTAL & 5552 & & & 165.511 \\
\hline
\hline
\end{tabular}
\end{center}}
\end{table}

\clearpage

\section{Observations and Data Reduction}

Observations of RZ LMi reported in present paper were obtained
during 46 nights between January 22, 2004 and May 28, 2005 at the
Ostrowik station of the Warsaw University Observatory and at CBA Concord
at the San Francisco suburb of Concord, approximately 50 km from East of
the City. The Ostrowik data were collected using the 60-cm Cassegrain
telescope equipped with a Tektronics TK512CB back-illuminated CCD
camera. The scale of the camera was 0.76"/pixel providing a
6.5'$\times$6.5' field of view. The full description of the telescope
and camera was given by Udalski and Pych (1992).

The Ostrowik data reductions were performed using a standard procedure
based on the IRAF\footnote{IRAF is distributed by the National Optical
Astronomy Observatory, which is operated by the Association of
Universities for Research in Astronomy, Inc., under a cooperative
agreement with the National Science Foundation.} package and
profile photometry was derived using the DAOphotII package
(Stetson 1987).

The CBA data were collected using an f/4.5 73-cm reflector operated at
prime focus on an English cradle mount. Images were collected with a
Genesis G16 camera using a KAF1602e chip giving a field of view of
$14.3'\times 9.5'$. Images were reduced using AIP4WIN software (Berry \&
Burnell 2000).

In both sites we monitored the star in ``white light'' in order to be
able to observe it with good precision also at minimum light of around
17 mag.

A full journal of our CCD observations of RZ LMi is given in Table 2. In
total, we monitored the star for 165.5 hours and obtained 5552
exposures.

Relative unfiltered magnitudes of RZ LMi were determined as the
difference between the magnitude of the variable and the intensity
averaged magnitude of two nearby comparison stars. The magnitudes and
colors of our comparison stars were taken from Henden and Honeycutt
(1995). Transformation to Johnson $V$ magnitudes was done using $BVR$
photometry of the field of variable obtained on 2004 Apr 21.

The accuracy of our measurements varied between 0.004 and 0.119 mag
depending on the brightness of the object and atmospheric conditions.
The median value of the photometric errors was 0.012 mag.

\section{General light curve}

The global light curve spanning whole period of our observations is
shown in Fig. 1. In total we detected 12 long eruptions and 7 short
outbursts. The superoutburst are labeled by corresponding roman numbers.
In quiescence the star fades to $V\approx 16.5$ mag and during the
highest phase of superoutburst reaches 13.9 mag giving the full
amplitude of variability equal to $A_s = 2.6$ mag. It is only slightly
larger than the value of 2.5 mag determined by Robertson et al. (1995).
During the brightest normal outburst the star reaches 14.4 mag.

First, from global light curve we selected only nights during which the
star was in superoutburst (it means that we detected clear superhumps).
Then we computed {\sc anova} statistics with two harmonic Fourier series
(Schwarzenberg-Czerny 1996). The resulting periodogram, for the
frequency range $0\div0.15$ c/d, is shown in Fig. 2. The dominant peak
is detected at frequency $f_0 = 0.05245(10)$ c/d, which corresponds to
the period of 19.07(4) days. This value is interpreted as supercycle
length i.e. mean interval between two successive superoutbursts. It is
in quite good agreement with value of 18.87 obtained by Robertson et al.
(1995). 

Next, we fitted analytical light curve to the superoutburst number V
(solid line in Fig. 1), which has very good coverage, and repeated it
every 19.07 days. The stability of the supercycle period is very
interesting. The analytical light curve has no problems with hitting 
precisely superoutbursts numbers I, II, III, VI and VII in 2004 and even
superoutbursts numbers XXIV and XXVI occurring one year later.

\vspace{10.6cm}

\includegraphics{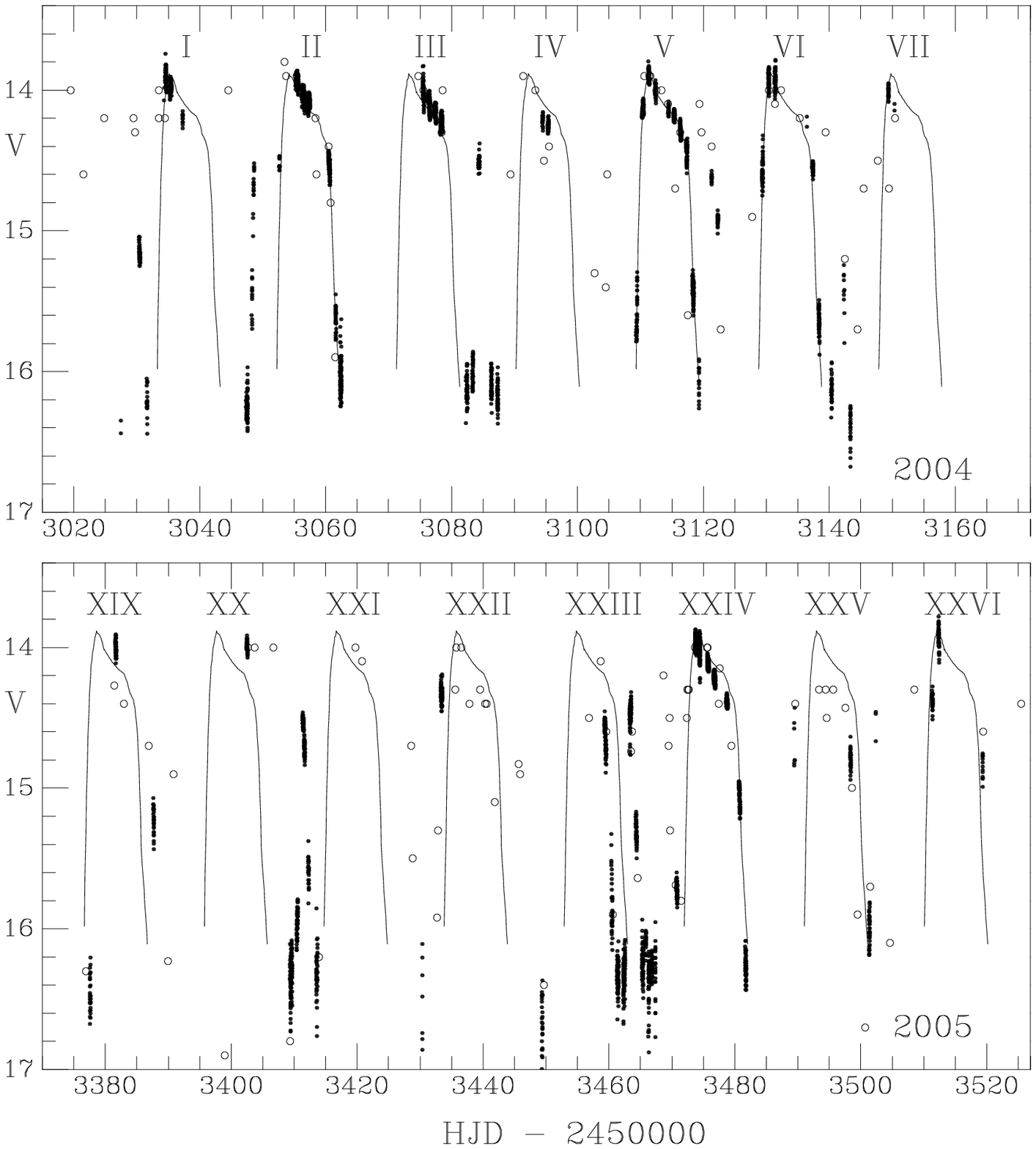}

   \begin{figure}[h]
      \caption{\sf The general photometric behavior of RZ LMi during
our campaign. Dots and open circles correspond to our and AAVSO observations.
The solid line fitted to eruption no. V is repeated every 19 days.}
   \end{figure}

\vspace{5.8cm}

\includegraphics{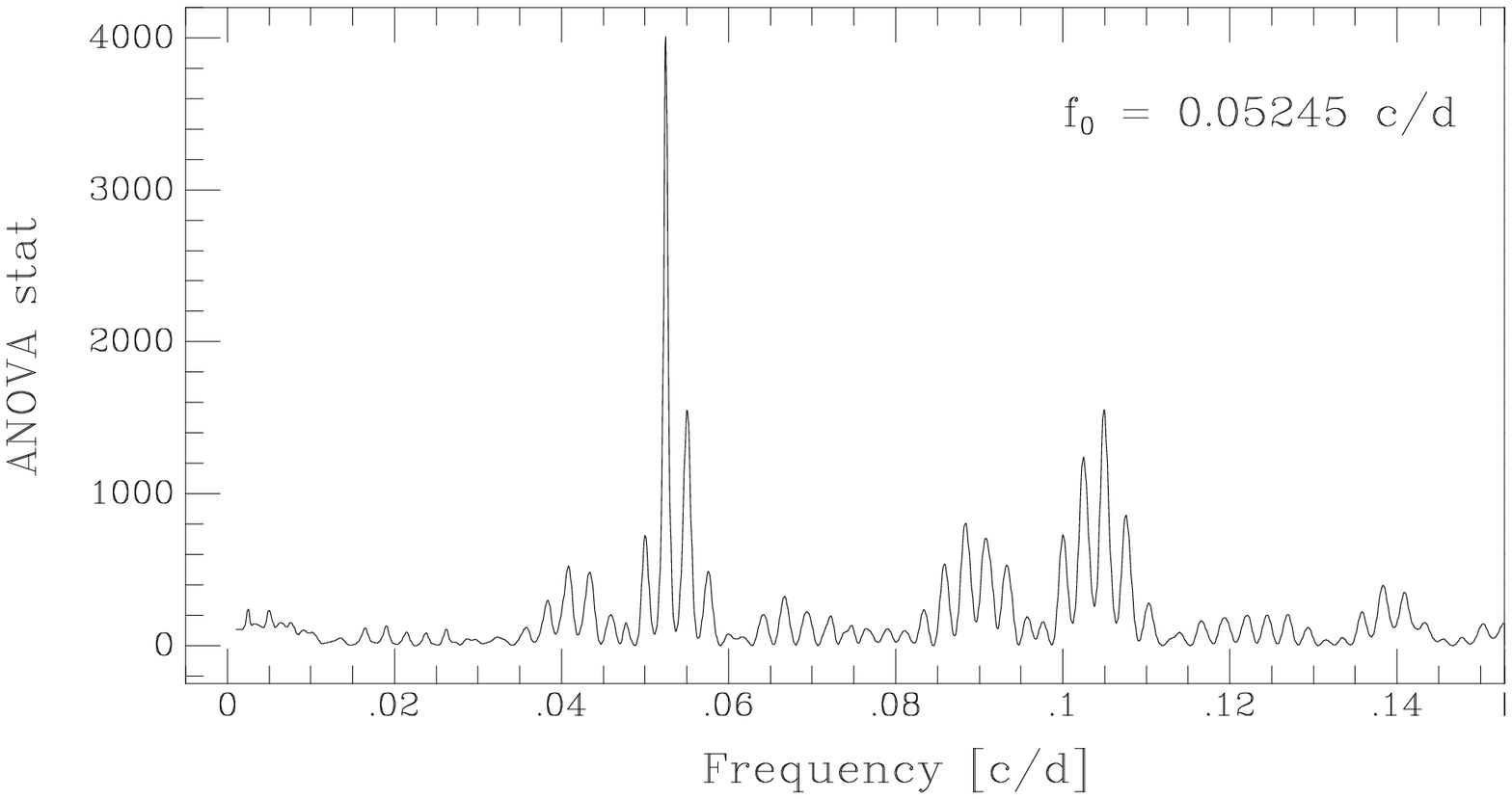}

   \begin{figure}[h]
      \caption{\sf The ANOVA spectrum of RZ LMi global light curve after
removing the data from quiescence and normal outbursts.
              }
   \end{figure}

Additionally, Fig. 3 shows the light curve consisting of only
superoutburst data and phased with period 19.07 days. One can clearly
see that superoutburst lasts slightly over half of the supercycle i.e.
over 10 days. It consists of: initial rise, which takes about 1.3 days,
plateau phase with linear decrease of brightness at rate of 0.063 mag/day
and lasting 7.5 days and final decline which takes about 1.5 days.

\vspace{5.5cm}

\includegraphics{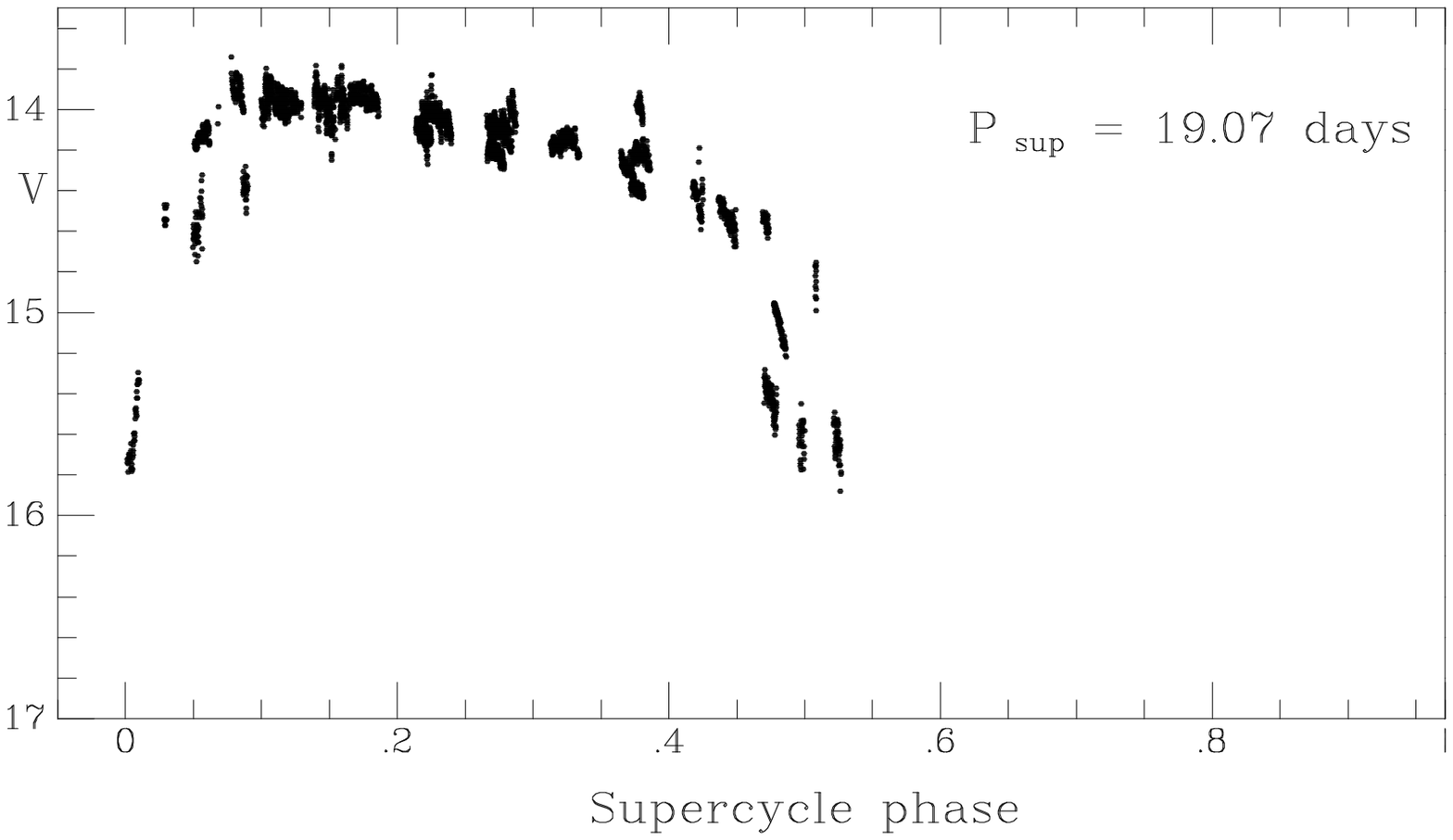}

   \begin{figure}[h]
      \caption{\sf The light curve of RZ LMi in superoutbursts obtained
by folding the general light curve with supercycle period of 19.07 days.
              }
   \end{figure}

\vspace{5.8cm}

\includegraphics{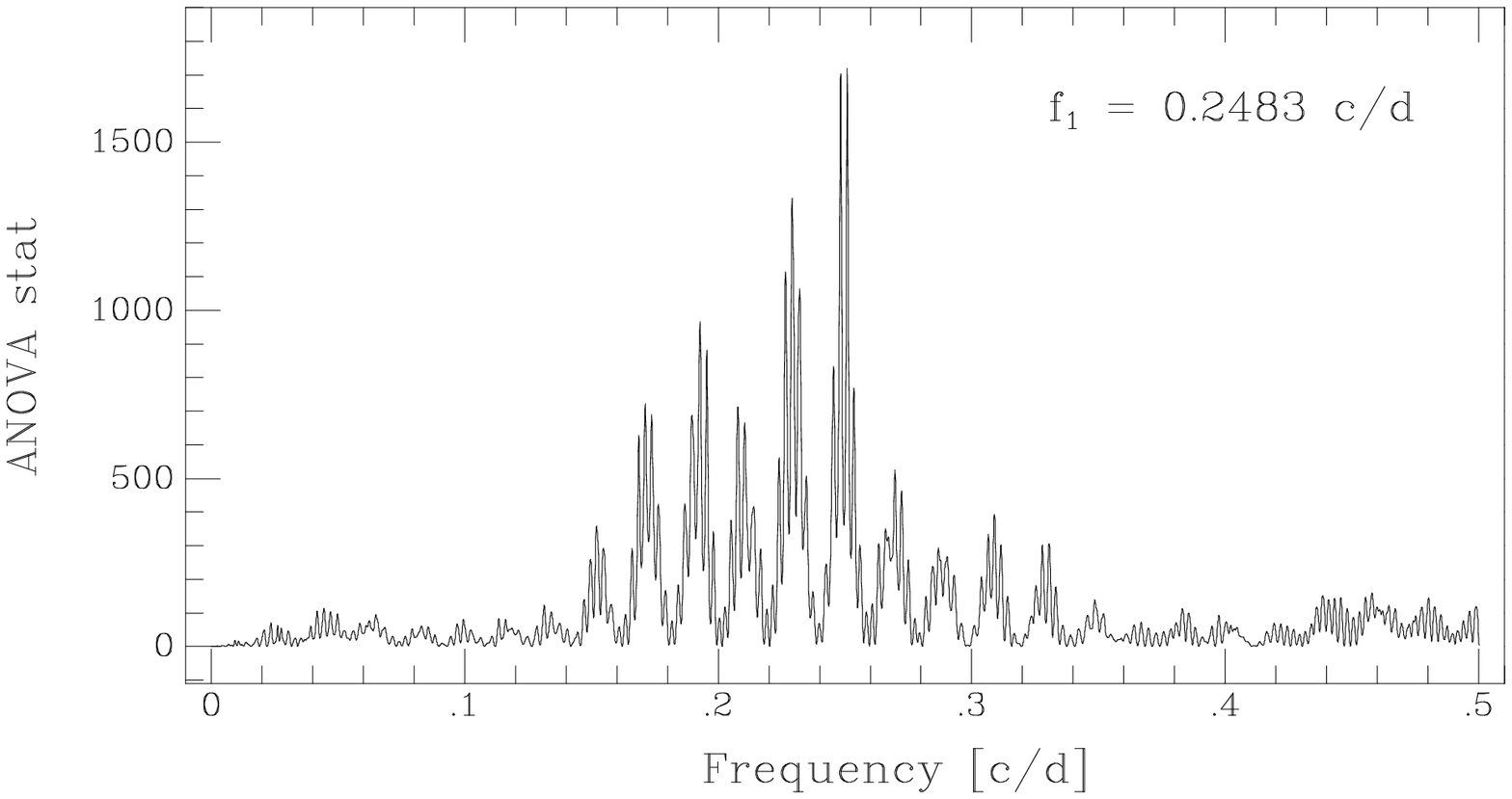}

   \begin{figure}[h]
      \caption{\sf The ANOVA spectrum of RZ LMi global light curve after
removing the data from superoutbursts.
              }
   \end{figure}

\vspace{5.5cm}

\includegraphics{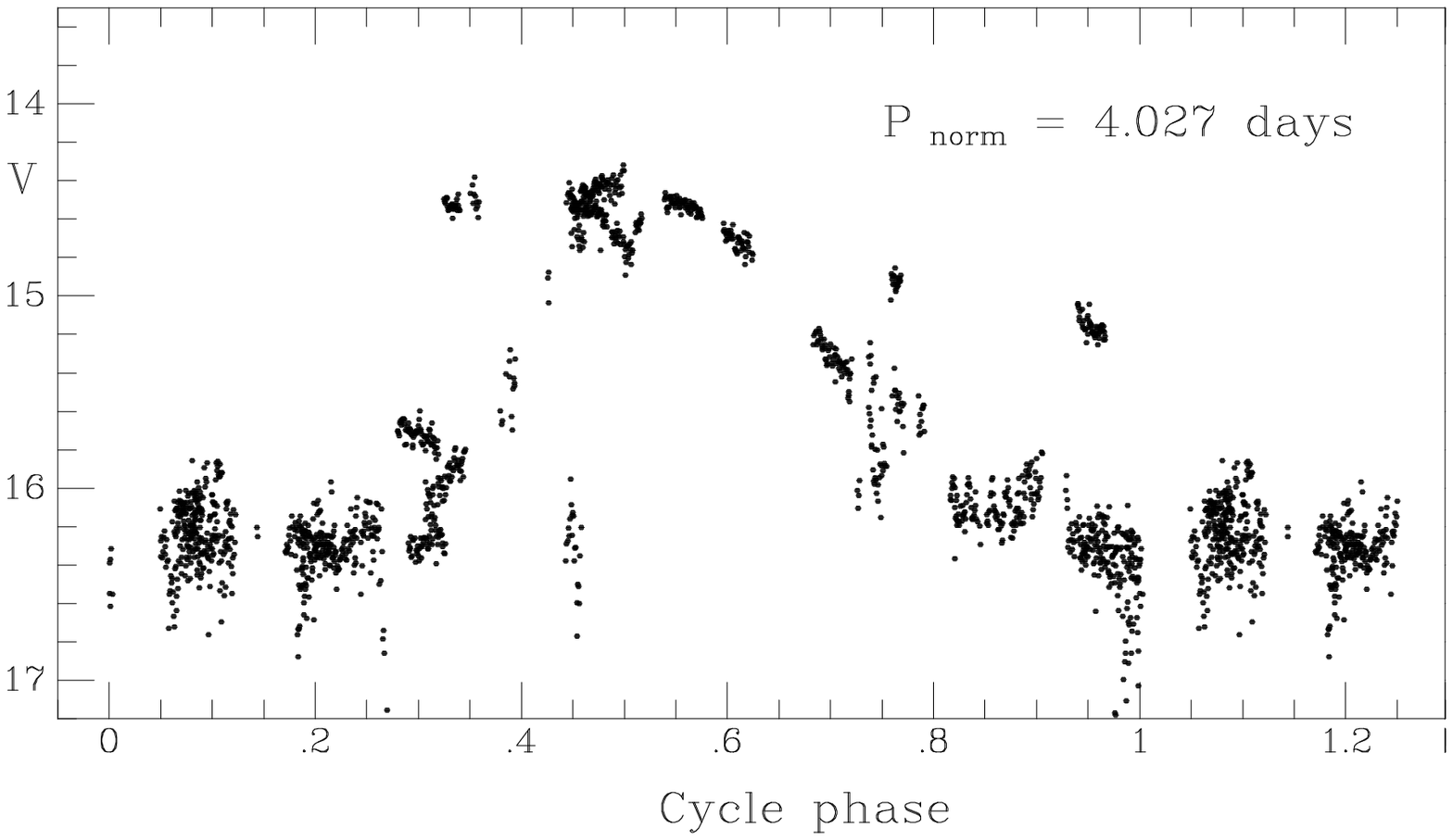}

   \begin{figure}[h]
      \caption{\sf The light curve of RZ LMi in normal outbursts and
quiescence obtained
by folding the general light curve with cycle period of 4.027 days.
              }
   \end{figure}

Now we can make opposite operation i.e. remove from the light curve all
superoutbursts and leave intervals when star is in quiescence and goes
into normal outbursts. Again, for the resulting light curve, we computed
the  {\sc anova} statistics and showed the result in Fig. 4. The
dominant peak has a double structure with maxima at frequencies
$f_1=0.2483(2)$ and $f_2=0.2509$ c/d. The phased light curve looks
better for the first frequency, and we choose it as correct value. The
corresponding period of $4.027(3)$ days is interpreted as normal cycle
i.e. interval between two successive normal outbursts. The light curve
phased with this period is shown in Fig. 5.

Normal outburst lasts 2.8 days and consists of quick initial rise lasting
only half a day, narrow maximum and slower decline. Taking into account the
fact that every supercycle we observe two normal outbursts, RZ LMi is in
the quiescence only for 3 days in each supercycle.

\section{Superhumps}

The superhumps of RZ LMi were observed on several occasions. Fig. 6 
shows data from three consecutive nights of superoutburst no. II 
which occurred in February 2004. Periodic, tooth-shape light variations
with amplitude of 0.1-0.2 mag are clearly visible.

\vspace{7.7cm}

\includegraphics{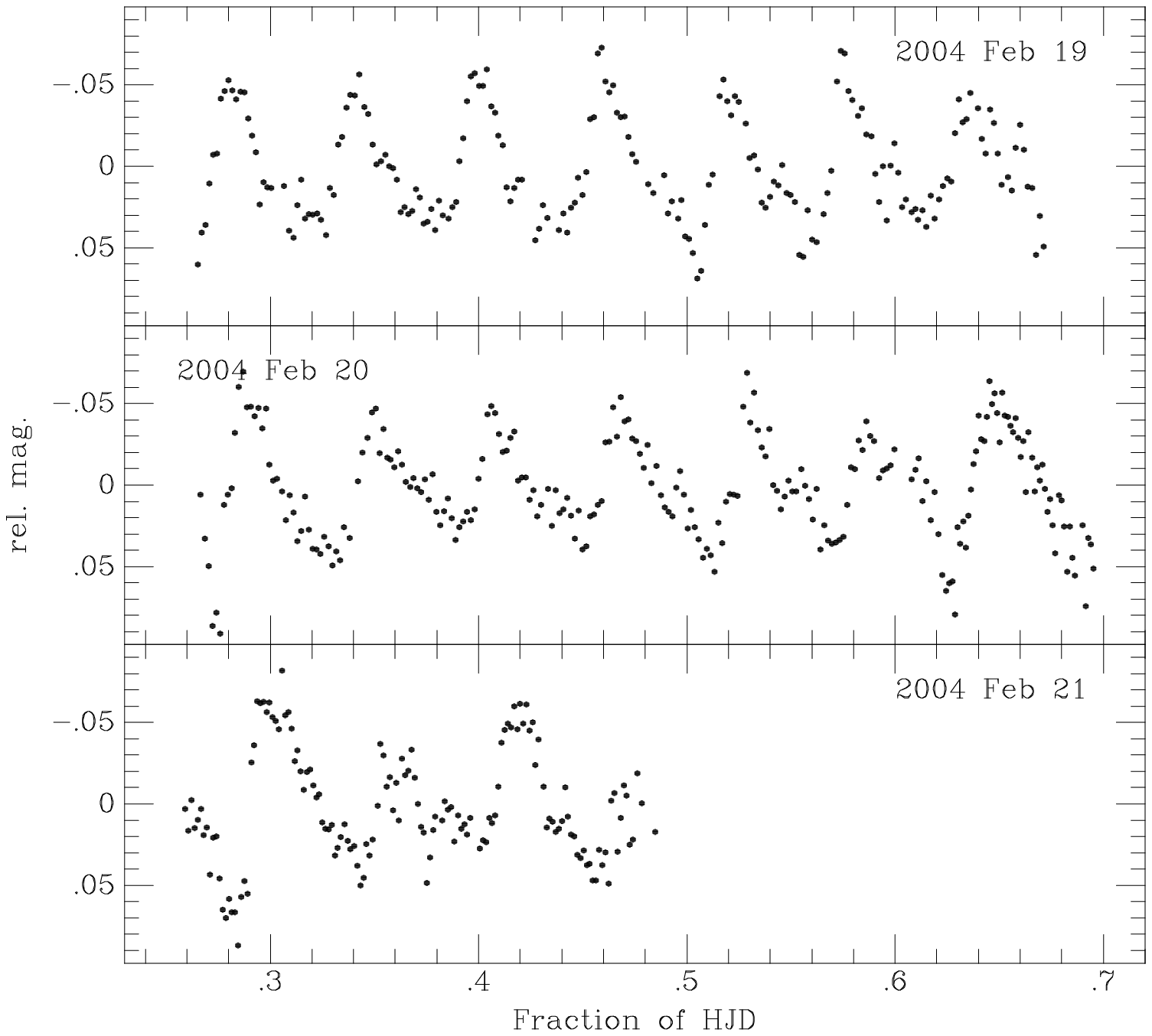}

   \begin{figure}[h]
      \caption{\sf Superhumps of RZ LMi from three consecutive nights
of February 2004.
              }
   \end{figure}

Additionally, Fig. 7 shows global light curve of superoutburst no. V,
which has the best observational coverage. The observing runs, due to
the geometric conditions, are not as long as in February, but the star
was observed on almost every night of the superoutburst. We were able to
see the initial rise (Apr 13), the birth of superhumps before the
maximum brightness (Apr 14), full amplitude variations which occurred one
night later, slow evolution towards smaller amplitudes occurring during
next five days of plateau phase and trace of superhumps during final
decline. 

\clearpage

~

\vspace{15.7cm}

\includegraphics{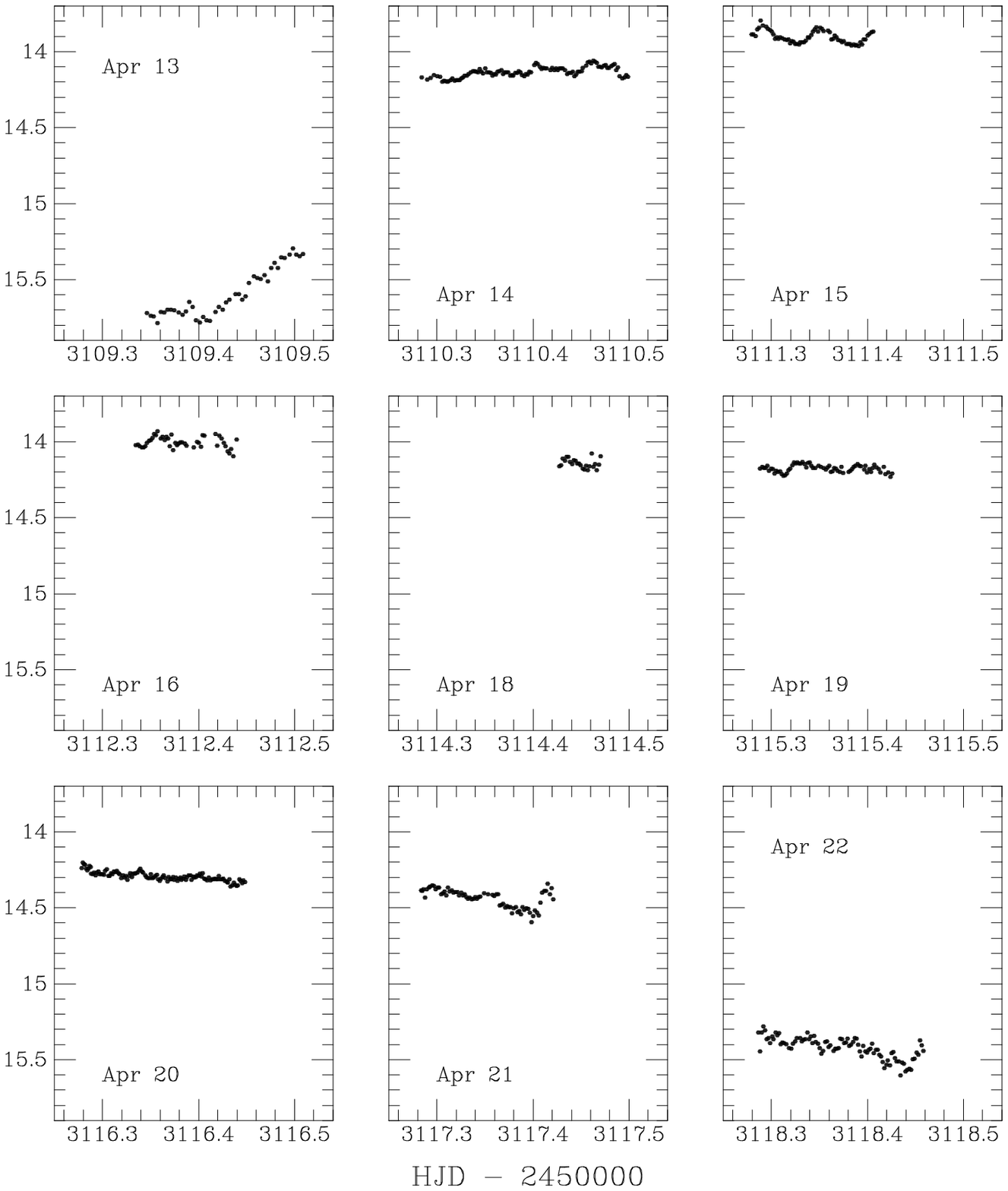}

   \begin{figure}[h]
      \caption{\sf Nightly light curves of RZ LMi from its April 2004
superoutburst.
              }
   \end{figure}

\subsection{{\sc anova} analysis}

The data from each night containing superhumps were fitted with straight
line or parabola. In purpose of detrending, this analytic curve was
subtracted from real light curve. As a result we obtained a  set of data
with average brightness equal to zero and consisting only short term
modulations.

For these sets we computed {\sc anova} statistics and showed corresponding
periodograms in Fig. 8. Additionally, the frequencies and periods determined
using these periodograms are summarized in Table 3.

\begin{table}[!h]
\caption{\sc Frequencies and periods of superhumps found in the 
periodograms computed
for detrended data of six superoutbursts.}
\vspace{0.1cm}
\begin{center}
\begin{tabular}{|l|l|c|c|}
\hline
\hline
Superoutburst & Date & $f_{\rm sh}$ [c/d] & $P_{\rm sh}$ [d]\\
\hline
No.   I & 2004, Jan 29 - Feb 01 & $16.8\pm0.1$     & 0.0595(4)\\
No.  II & 2004, Feb 19 - Feb 24 & $16.824\pm0.020$ & 0.05944(7)\\
No. III & 2004, Mar 10 - Mar 13 & $16.831\pm0.025$ & 0.05941(9)\\
No.   V & 2004, Apr 14 - Apr 21 & $16.823\pm0.010$ & 0.05944(4)\\
No.  VI & 2004, May 04 - May 12 & $16.828\pm0.020$ & 0.05942(7)\\
No. XXIV& 2005, Apr 05 - Apr 18 & $16.836\pm0.025$ & 0.05940(9)\\
\hline
Mean    & 2004 - 2005           & $16.8363\pm0.001$ & 0.059396(4) \\
\hline
\hline
\end{tabular}
\end{center}
\end{table}

The main frequencies detected in each superoutburst are consistent
within the errors with each other and power spectrum computed for all
superoutbursts returns the mean frequency $f_{\rm sh}=16.8363\pm0.001$,
which corresponds to the period of $P_{\rm sh}=0.059396(4)$ days
($85.530\pm0.006$ min), confirming that RZ LMi is one of the shortest
period SU UMa, and particularly ER UMa, stars. 

\vspace{13.7cm}

\includegraphics{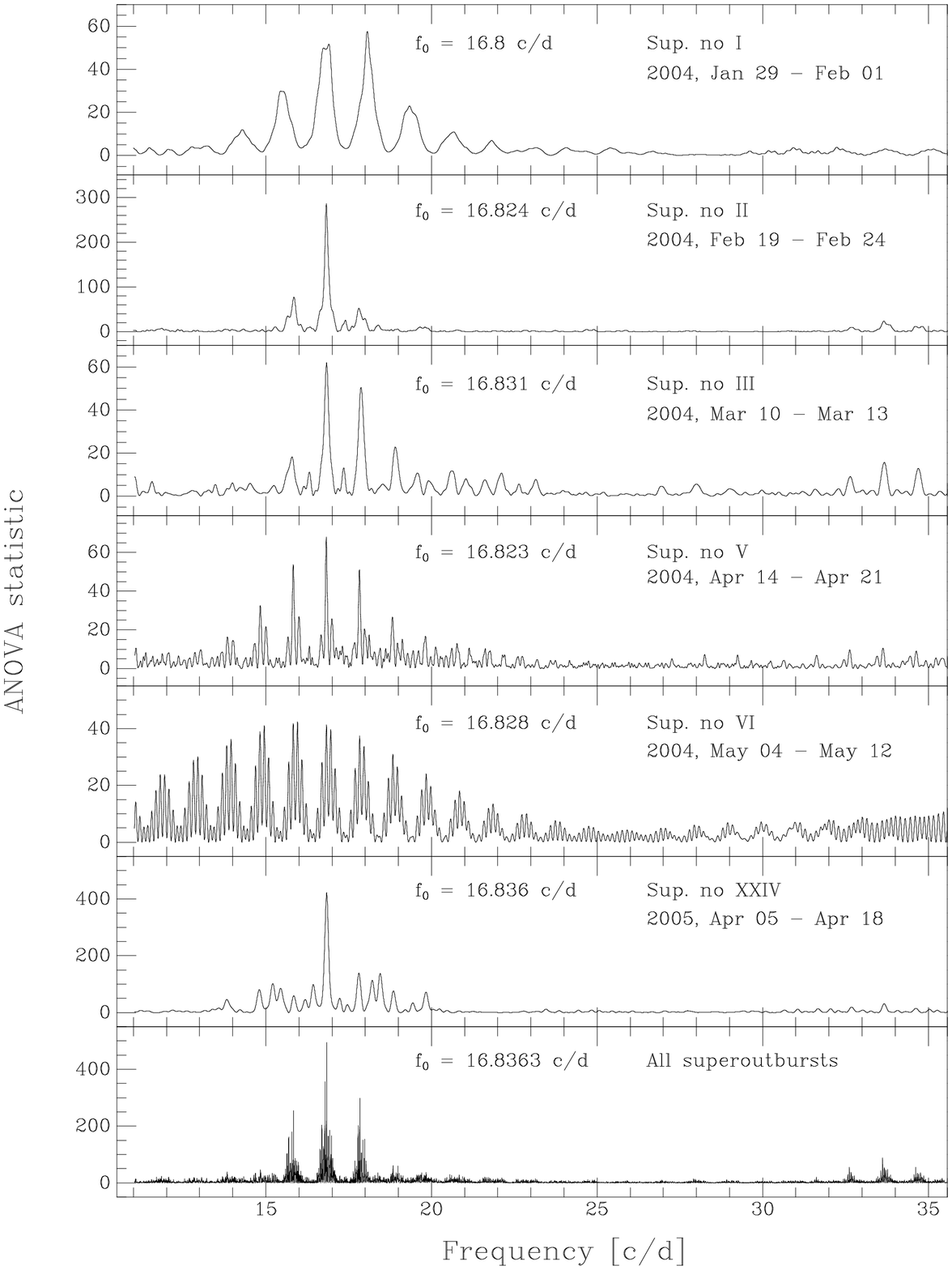}

   \begin{figure}[h]
      \caption{\sf ANOVA power spectra for superhumps observed in
six superoutbursts of RZ LMi and composite spectrum obtained from
all data from supermaxima.
              }
   \end{figure}

\begin{table}[!h]
\caption{\sc Cycle number $E$, $O-C$ values and times of maxima for
superhumps observed in six superoutbursts. Note that the first three
superoutbursts have common $E$ numbering.}
\vspace{0.1cm}
\begin{center}
\begin{tabular}{|r|c|r|r||r|c|r|r|}
\hline
\hline
$E$ & ${\rm HJD}_{\rm max} - 2453000$ & Error & $O-C$ & 
$E$ & ${\rm HJD}_{\rm max} - 2453000$ & Error & $O-C$\\
\hline
0&34.6080&0.0040&0.0069&0&110.3440&0.0030&0.0316\\
1&34.6640&0.0050&-0.0504&1&110.4033&0.0020&0.0302\\
13&35.3820&0.0020&0.0362&2&110.4610&0.0025&0.0018\\
14&35.4372&0.0025&-0.0346&16&111.2905&0.0025&-0.0303\\
\cline{1-4}
348&55.2830&0.0035&0.0422&17&111.3505&0.0027&-0.0199\\
349&55.3418&0.0025&0.0320&34&112.3573&0.0025&-0.0665\\
350&55.3992&0.0030&-0.0018&35&112.4147&0.0035&-0.0999\\
351&55.4592&0.0020&0.0082&69&114.4360&0.0025&-0.0633\\
352&55.5198&0.0025&0.0284&84&115.3290&0.0030&-0.0262\\
353&55.5745&0.0020&-0.0508&85&115.3895&0.0025&-0.0074\\
354&55.6390&0.0030&0.0349&101&116.3390&0.0025&-0.0188\\
365&56.2900&0.0030&-0.0064&102&116.4010&0.0030&0.0252\\
366&56.3500&0.0023&0.0036&117&117.2958&0.0025&0.0927\\
367&56.4065&0.0025&-0.0453&118&117.3545&0.0030&0.0812\\
368&56.4680&0.0030&-0.0100&119&117.4155&0.0030&0.1083\\
\cline{5-8}
369&56.5300&0.0025&0.0337&337&130.3560&0.0020&0.0129\\
370&56.5870&0.0030&-0.0068&338&130.4170&0.0030&0.0401\\
371&56.6465&0.0025&-0.0052&354&131.3630&0.0020&-0.0303\\
382&57.3000&0.0030&-0.0044&355&131.4207&0.0030&-0.0587\\
\cline{5-8}
383&57.3600&0.0030&0.0056&0&473.7042&0.0020&-0.0049\\
384&57.4200&0.0025&0.0156&1&473.7638&0.0020&-0.0018\\
436&60.5060&0.0035&-0.0358&2&473.8230&0.0015&-0.0054\\
438&60.6235&0.0035&-0.0579&10&474.3040&0.0035&0.0900\\
455&61.6340&0.0035&-0.0475&11&474.3630&0.0030&0.0830\\
\cline{1-4}
702&76.3160&0.0035&0.1038&12&474.4220&0.0030&0.0760\\
705&76.4910&0.0035&0.0497&33&475.6640&0.0025&-0.0207\\
706&76.5450&0.0020&-0.0413&34&475.7228&0.0015&-0.0311\\
719&77.3210&0.0025&0.0216&35&475.7825&0.0025&-0.0263\\
720&77.3830&0.0030&0.0653&36&475.8440&0.0020&0.0088\\
721&77.4410&0.0025&0.0417&50&476.6750&0.0025&-0.0052\\
722&77.4965&0.0030&-0.0241&51&476.7325&0.0030&-0.0374\\
737&78.3905&0.0030&0.0252&52&476.7924&0.0027&-0.0293\\
738&78.4485&0.0020&0.0016&53&476.8515&0.0022&-0.0346\\
739&78.5032&0.0023&-0.0776&84&478.6983&0.0030&0.0478\\
   &       &      &          &85&478.7580&0.0030&0.0525\\
   &       &      &          &86&478.8160&0.0030&0.0287\\
\hline
\hline
\end{tabular}
\end{center}
\end{table}

\subsection{The $O - C$ analysis}

In the light curve of RZ LMi from all superoutbursts we detected 70
maxima of superhumps. Their times are listed in Table 4 together with
the errors, cycle number $E$ and $O-C$ values computed according to the
ephemeris which will be described further.

The $O-C$ values from first three superoutbursts shows no signs
of significant trend indicating that the period of superhumps was
roughly constant. 

There are observational evidences that ER UMa stars shows ordinary
superhumps also in quiescence indicating that in these systems the
disk is elliptical and tidally unstable all the time. It might suggest
that the star should remember the phase of the superhumps from one
superoutburst to another. The $O-C$ data from our superoutbursts
number I, II and III seem to confirm this hypothesis. They can be
fitted with common linear ephemeris in the form:

\begin{equation}
{\rm HJD_{\rm max}} = 2453034.6076(10) + 0.059405(2) \cdot E
\end{equation}

The corresponding $O-C$ diagram is shown in Fig. 9.

\vspace{7cm}

\includegraphics{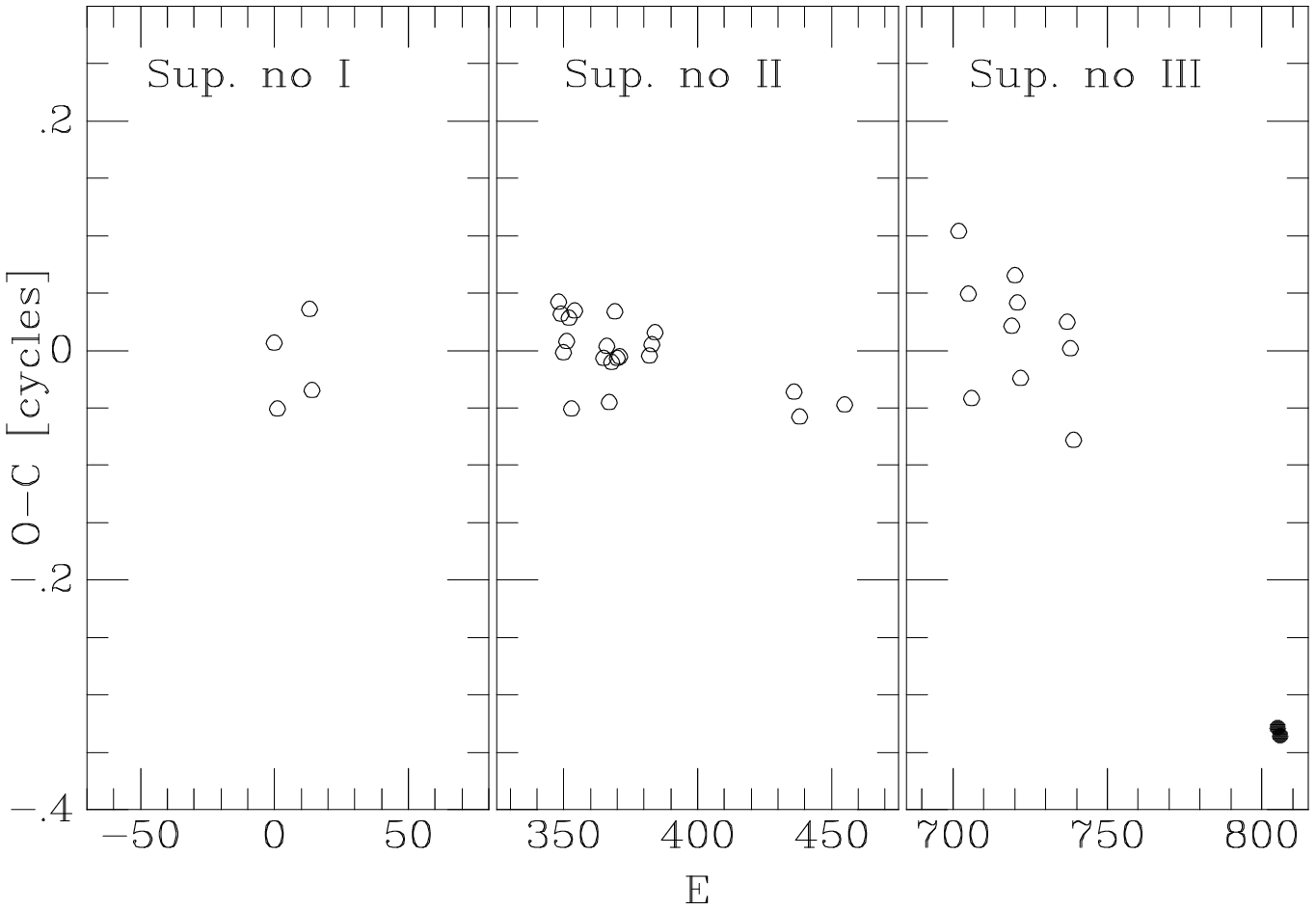}

   \begin{figure}[h]
      \caption{\sf The $O-C$ diagram for superhumps maxima of RZ LMi
detected during its superoutbursts number I, II and III. Black dots
correspond to possible late superhumps described in Sect. 6.
              }
   \end{figure}

Moreover, the detrended light curve containing superhumps from all
superoutbursts might be phased with one period and shows no traces
of phase shifts between superhumps from different superoutburst. Such
a light curve is plotted in Fig. 10.

\vspace{5cm}

\includegraphics{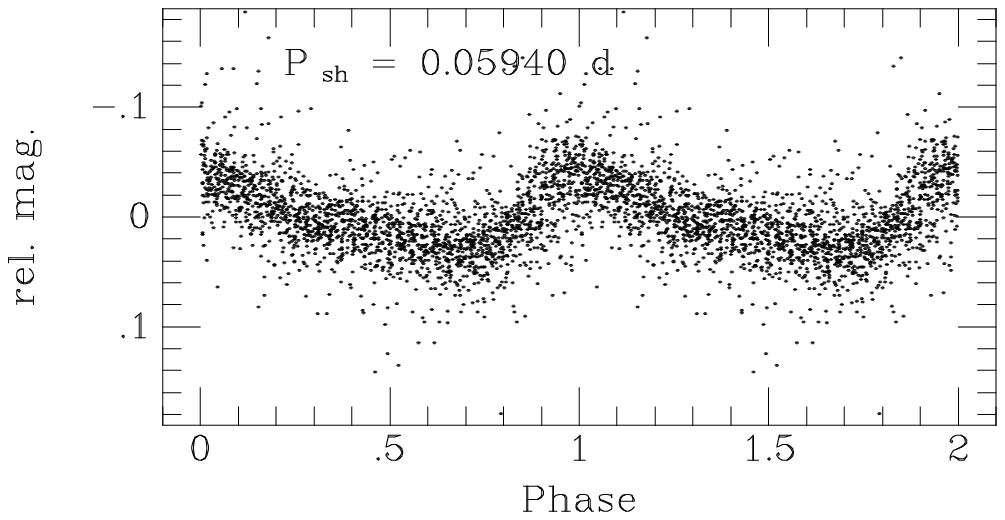}

   \begin{figure}[h]
      \caption{\sf The detrended light curve from data collected
during all superoutbursts observed in 2004 folded on superhump
period.
              }
   \end{figure}

Something strange happened to RZ LMi during superoutburst number
IV. We detected then a clear eruption, which has properies of superoutburst
i.e. is brighter that ordinary outburst and shows decline typical for
plateau phase but during two nights of this bright state we have not
detected any superhumps.

The superoutburst no. V occurred in right time but with slightly
different behaviour of superhumps. Their maxima can be fitted with
following linear ephemeris:

\begin{equation}
{\rm HJD_{\rm max}} = 2453110.3408(11) + 0.059414(15) \cdot E
\end{equation}

\noindent but from the $O-C$ values computed according to this ephemeris
and shown in Table 4 and in Fig. 11 it is evident that the period of
superhumps was quickly increasing. Thus the moments of maxima can be
fitted with the following parabola:

\begin{equation}
{\rm HJD_{\rm max}} = 2453110.3436(13) + 0.059152(65) \cdot E + 2.27(55)\cdot
10^{-6}\cdot E
\end{equation}

\noindent indicating that the period was increasing with the rate of
$\dot P/P_{\rm sh} = 7.6(1.9)\cdot 10^{-5}$. Such a period derivative is
typical for SU UMa stars with superhump periods of around 0.06 days
(for example see Fig. 5 in Rutkowski et al. 2007).

\vspace{5.5cm}

\includegraphics{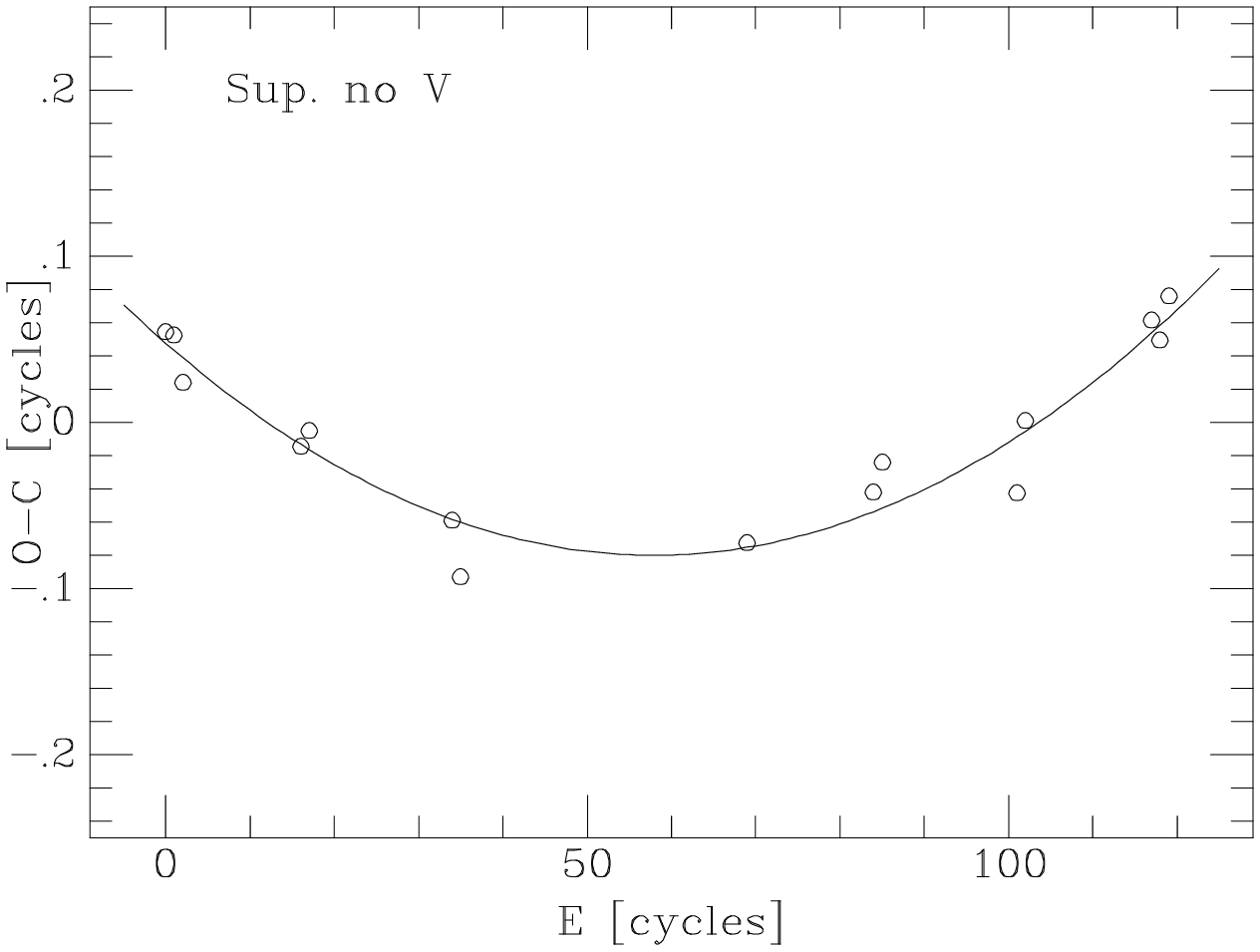}

   \begin{figure}[h]
      \caption{\sf The $O-C$ diagram for superhumps maxima of RZ LMi
detected during its superoutburst number V.
              }
   \end{figure}

\vspace{5.5cm}

\includegraphics{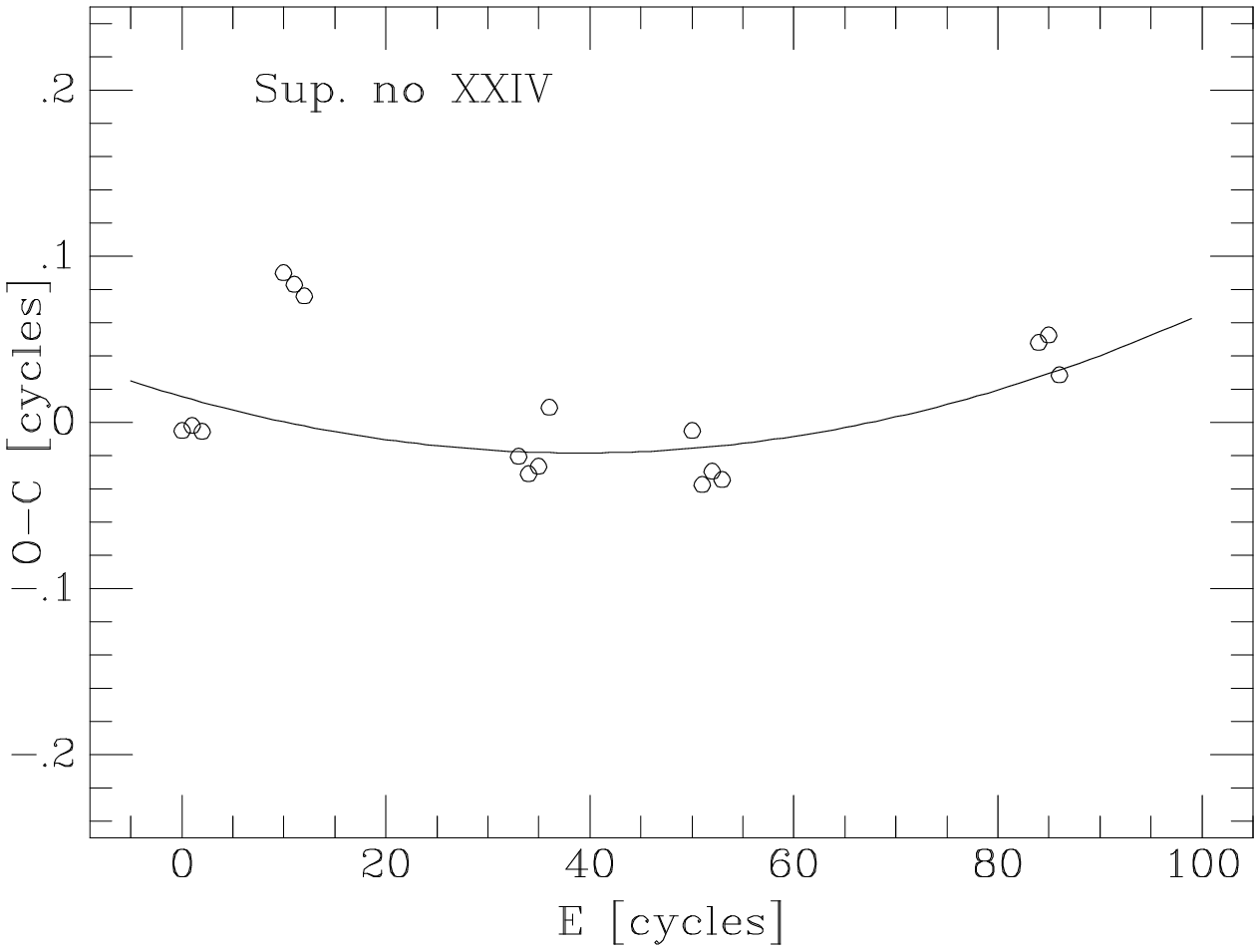}

   \begin{figure}[h]
      \caption{\sf The $O-C$ diagram for superhumps maxima of RZ LMi
detected during its superoutburst number XXIV.
              }
   \end{figure}

There are insufficient number of data to investigate possible period
changes during superoutburst no. VI, thus the corresponding moments
of maxima were fitted only with the linear ephemeris:

\begin{equation}
{\rm HJD_{\rm max}} = 2453130.3566(17) + 0.05918(14) \cdot E
\end{equation}

There was only one superoutburst with sufficient amount of data for
$O-C$ analysis in 2005 season. It was superoutburst no. XXIV and its
maxima can be fitted with the following linear ephemeris:

\begin{equation}
{\rm HJD_{\rm max}} = 2453473.7045(9) + 0.059416(21) \cdot E
\end{equation}

However, the data collected in Table 4 and shown in Fig. 12 might 
suggest slight increasing trend with rate of $\dot P/P_{\rm sh} =
4.5(2.5)\cdot 10^{-5}$. On the other hand, the error of this determination
is large, and within $2 \sigma$ it is consistent with constant value
of period.

\section{Quiescence and normal outbursts}

As we wrote earlier RZ LMi is so active that it is difficult to find
it in quiescence. However, on three occasions, we collected sufficient
amount of data to make the analysis of behaviour of the star in minimum
light and in ordinary outbursts.

\vspace{9.5cm}

\includegraphics{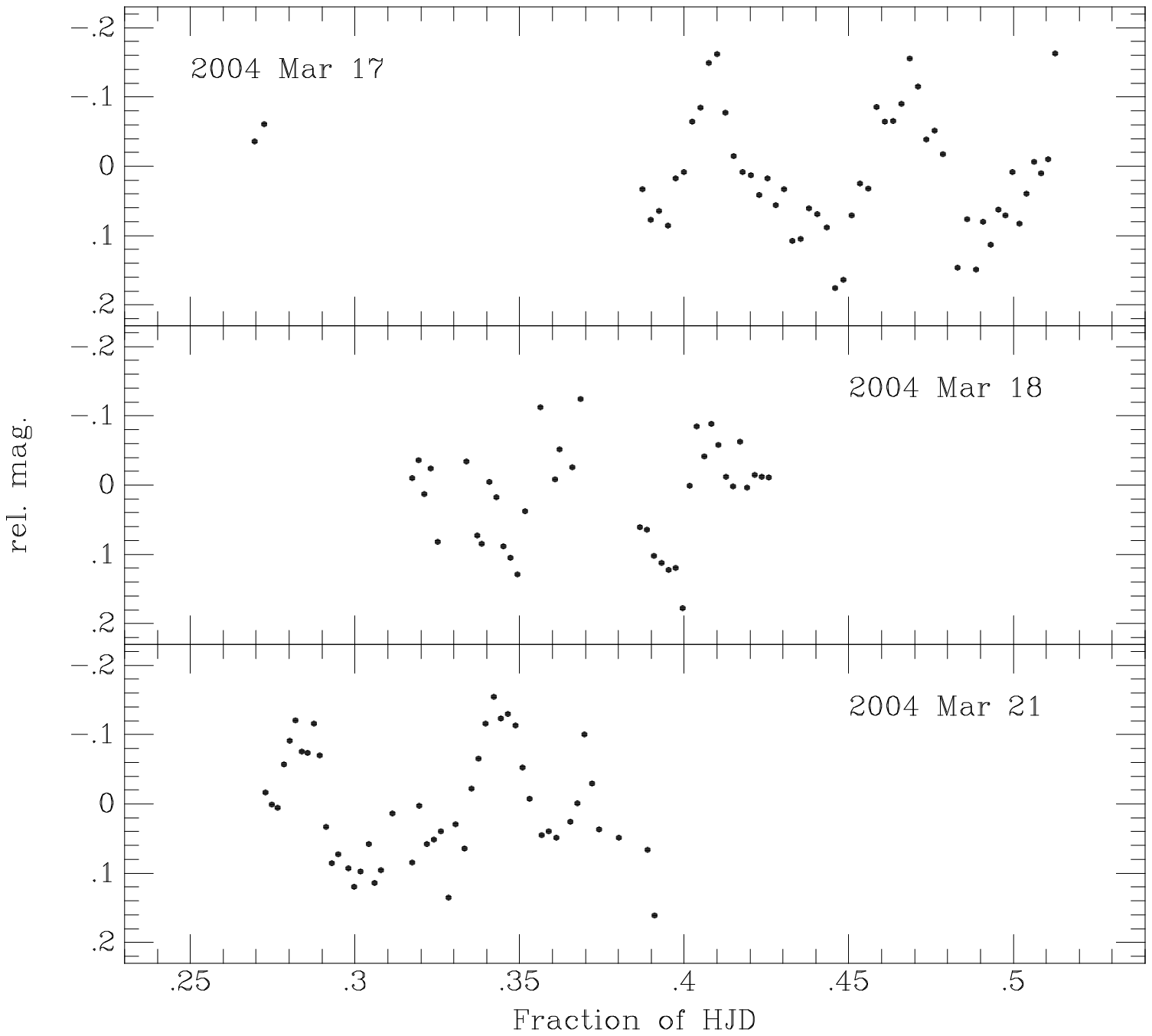}

   \begin{figure}[h]
      \caption{\sf Sample light curves of RZ LMi from quiescence.
              }
   \end{figure}

The first interval of data comes from 2004, Mar 17-22 when we observed
RZ LMi on four nights of minimum light and one night of the normal 
outburst. Sample light curves from these period are shown in Fig. 13 and
display clear and periodic light variations of amplitude around 0.3-0.4
mag. Taking into account that these data were collected just after the
final decline of superoutburst no. III, one can suspect that we observe
so called late superhumps - the phenomenon occurring at the end of
superoutburst with period roughly equal to period of ordinary superhumps
but with phase shift reaching up to 0.5 cycle. $O-C$ diagram from Fig.
9 shows the moments of the maxima observed on 2004 Mar 17 as black dots
suggesting that they are shifted in phase by about 0.3 cycle i.e.
significantly less than typical value of 0.5 cycle.

\vspace{8.5cm}

\includegraphics{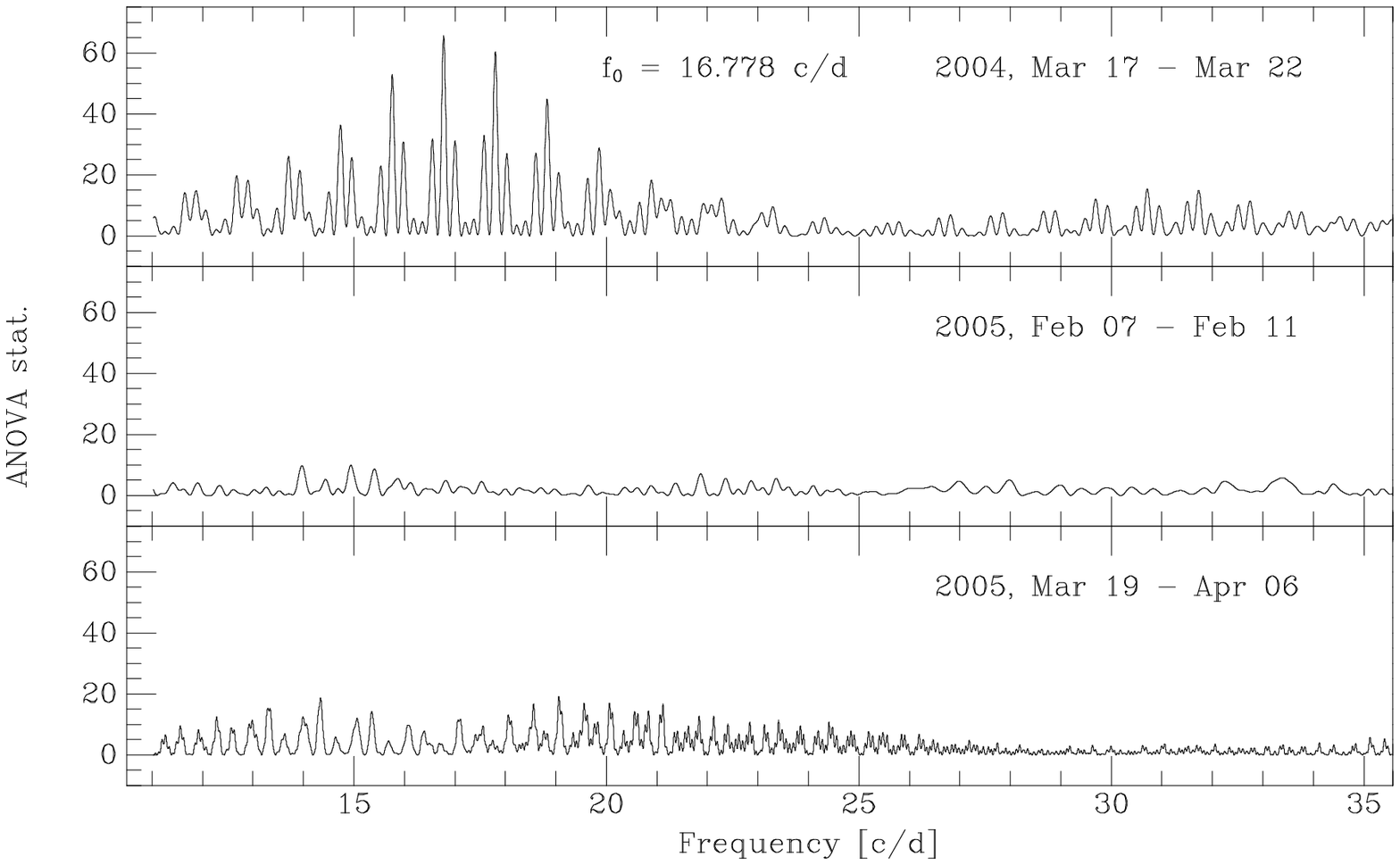}

   \begin{figure}[h]
      \caption{\sf ANOVA power spectra for three long runs covering
the quiescence and normal outbursts.
              }
   \end{figure}

To find a period of these variations, we first transformed our light
curves to the intensity units, next we detrended them removing long
scale behaviour. The resulting {\sc anova} periodogram is shown in upper
panel of Fig. 14. The highest peak occurs at frequency
$f_0=16.778\pm0.02$ c/d corresponding to the period of 0.05960(7) days.
This is only 0.3\% longer than mean superhump period and the two
periods differ by the value which is about three times larger that the
error of the period determination. It is also possible that true value
of frequency appears as 1-day alias at $f_0=17.778\pm0.02$ c/d
corresponding to the period of 0.05625(7) days which is significantly
shorter than superhump period and might be also shorter than unknown
orbital period of the system. In this case this period might be assumed
as period of negative superhumps. However, it is known that negative
superhump, orbital and positive superhump periods correlate with each
other (Retter et al. 2002, Olech et al. 2007). This correlation
indicates that the orbital period should be around 0.0574 days and
superhump period excess $\epsilon$ should be as large as 3.5\% i.e.
about three times too high for star with such a superhump period.

Thus the final conclusion is that in quiescence RZ LMi showed
modulations with period roughly equal to superhump period and indicating
that in this interval the disc could be still eccentric and precessing.

Two other long intervals when the star was observed in quiescence
occurred on 2005, Feb 07 - 11 and 2005, Mar 19 - Apr 06. From two lower
periodograms shown in Fig. 14 it is clear that no periodic modulations
were observed at that time.

\section{Discussion}

\subsection{Evolutionary status of RZ LMi}

From our Table 1 summarizing main properies of ER UMa stars, it is clear
that these objects have many common properties but may be divided into
two subgroups probably with different evolutionary status. Fig. 15,
repeated after Patterson (1998, 2001) and Olech et al. (2004), shows
correlation between period excess (i.e. mass ratio) and orbital period
of the system. The solid line shows the evolutionary track of a dwarf
nova with a white dwarf of mass 0.75 $\cal M_\odot$ and secondary
component with effective radius 6\% larger than that of single main
sequence star. The nova evolves towards the shorter periods first due to
the magnetic braking, next due to the emission of gravitational waves.
After reaching the period minimum, the secondary becomes degenerate
brown dwarf and system starts to increase its orbital period.

\vspace{10.2cm}

\includegraphics{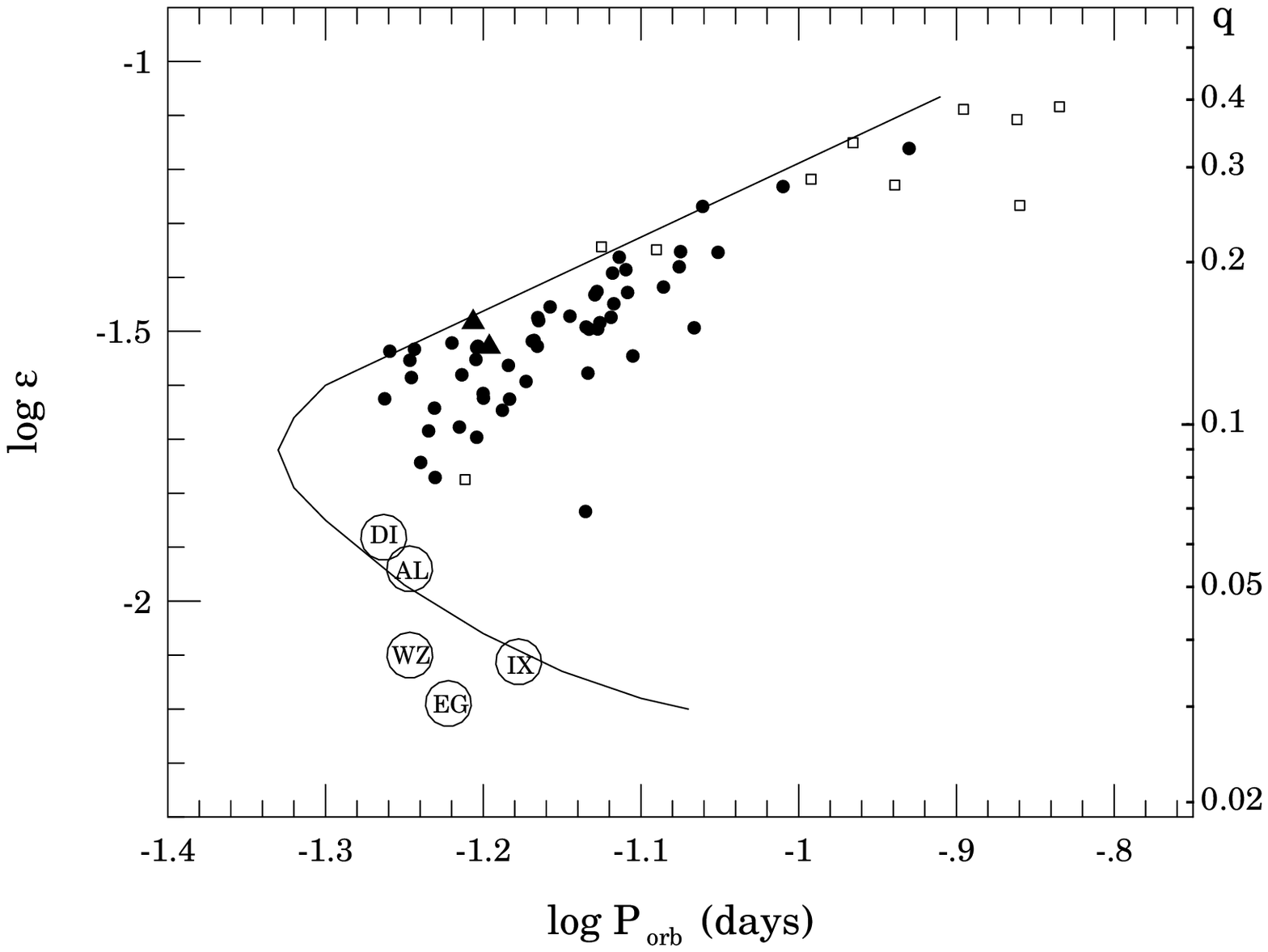}

   \begin{figure}[h]
      \caption {\sf The relation between the period excess and
orbital period of the system. The solid line corresponds to the
evolutionary track of a binary with a white dwarf of $0.75 {\cal M}_\odot$
and a secondary with effective radius 6\% larger than in the case of an
ordinary main sequence star. Calculations were made under the
assumption that below the orbital period of two hours the angular
momentum loss in only due to gravitational radiation. Triangles
denote the positions of ER UMa and V1159 Ori.}
\end{figure}

It seems that DI UMa and IX Dra (both belonging to ER UMa stars) are
such evolved period bouncers, which in fact should be similar to old
and inactive WZ Sge stars (WZ Sge, AL Com and EG Cnc showed in the
plot). On the other hand, ER UMa and V1159 Ori, shown as filled
triangles, seem to be much younger objects still evolving towards
shorter periods.

Where is the place of RZ LMi? It is difficult to answer this question
without knowledge about the orbital period of the system. Our
photometric data showed no other short term modulations than these
corresponding to the ordinary superhumps. It would be very
tempting to make the spectroscopic observations of the star in
quiescence. With minimum brightness of 16.5 mag it can be done with
2-3-meter class telescope.

\subsection{Stability of the supercycle}

The comprehensive analysis of the global light curve of RZ LMi made by
Robertson et al. (1995) and  based on almost three years observing
period showed that supercycle of RZ LMi is not stable. Their $O-C$
diagram for supermaxima was characterized by clear decreasing trend with
$\dot P = -1.7 \cdot 10^{-3}$. However graph shows also occasional jumps
where particular superoutburst occur even 5 days before or after the
predicted moment. If this decreasing trend would continue to the epoch
of our observations the supercycle should be then around 18.5 days,
which is in disagreement with determined value of 19.07 days.

Our global light curve spans only two seasons and has no enough data to
construct reliable $O-C$ diagram for supermaxima. However, quick look at
Fig. 1, could draw some valuable conclusions. In 2004 the 19-day
periodicity is preserved through all superoutbursts except eruption
number IV. In this case, we, in fact, are not certain whether we deal
with superoutburst which occurred slightly before predicted moment or
exceptionaly bright normal outburst lasting longer than usual. Vicinity
of eruption number IV is also the time when disk could loose its
eccentricity, expel the matter via this long outburst and rebuilt
eccentricity again in superoutburst no. V.

Data from 2005 seem to confirm stability of 19-day supercycle. The
superoutburst no. XXIV, which has the best observational coverage, occurs
at right time according to 19-day ephemeris. The problem is  with
superoutburst no. XXIII, where instead of supermaximum we noted two
ordinary outbursts. Our light curve, however, does not exclude
possibility that supermaximum occurred a few days earlier according to
the ephemeris.

Mass transfer from the secondary to the disk, building the
eccentricity, ignition of the outbursts and superoutbursts due to the
thermal and tidal instabilities are stochastic processes, which are far
for regularity. The question is why RZ LMi is so regular? Even if we
observe some shifts in time of the start of particular supermaximum, the
clock returns to stability without shift of the phase of whole pattern.
This is hard to explain from the point of view of standard thermal-tidal
instability model and might need some help from, for example, external
force. The present number of known SU UMa systems reached the level for
which the statistics tells us that some of these close binaries might be
orbited by a third body. Is this in case of RZ LMi? We do not know. But
the hypothesis that 19-day period is the orbital period of the third
body (or some kind resonant value) and cause of both the stability of
supercycle and high activity of the star, which without this body would
be quiet WZ Sge object, is tempting.

\subsection{Permanent superhumper?}

The standard thermal-tidal instability model is unable to produce
supercycles shorter that 40 days. Activity of the ordinary SU UMa
variable can be increased by increasing a mass transfer rate. But when
it reaches $\dot M \approx 3\cdot 10^{16}$ g/s the supercycle starts to
lenghten again due to the fact that superoutburst lasts longer. Further
increasing of mass transfer causes transition of the star to the group
of permanent superhumpers which are in permanent state of supermaximum
and show infinite value of supercycle.

Osaki (1995) tried to explain properties of RZ LMi by artificial ending
the superoutburst at the moment, when the disk had shrunk from $0.46a$
to only $0.42a$, whereas a typical value used for ordinary SU UMa
stars is $0.35a$.

Hellier (2001) suggested that the source of the premature end of
superoutburst might be a cooling wave propagating from the region
outside $0.46a$ causing transition of the disk to the cold state when
still eccentric. This decoupling of tidal and thermal stability brings
the star to the minimum light with still precessing and elliptical disk.
This hypothesis is confirmed by observations of ordinary superhumps
both in quiescence and normal outbursts of V1159 Ori and ER UMa
(Patterson et al. 1995, Gao et al. 1999, Zhao et al. 2006).

Our observations shows that RZ LMi also shows superhumps in minimum
light. Additionally, for the first time, we demonstrated that in interval
covering at least 60 days (including superoutbursts numbers I, II and
III) the star was showing superhumps with constant period which can be
described by common ephemeris and phased without any phase shift. It
indicates that decoupling could have place in this case and the disk of
RZ LMi was eccentric and precessing in the entire 60-day period.

\section{Summary}

We have presented the results of two seasons observational campaign 
devoted to RZ LMi. In total we detected 12 superoutbursts and 7 normal
outbursts. Our main findings may be summarized as follows:

\begin{itemize}

\item The $V$ brightness of the star varies in range from 16.5 to 13.9
mag. The superoutbursts occur every 19.07(4) days and last slightly over
10 days. The interval between two successive normal outbursts is 4.027(3)
days.

\item The mean period of superhumps observed during all superoutbursts
is $P_{\rm sh}=0.059396(4)$ days ($85.530\pm0.006$ min).

\item During three consecutive superoutbursts of 2004 the superhump
period was constant and the star "remembered" the phase of the superhumps
from one superoutburst to another. It supports the hypothesis that ER UMa
stars have accretion disks which are tidally unstable over long periods
of time.

\item The period of superhumps detected in superoutburst no. V was
increasing with the rate of $\dot P/P_{\rm sh} = 7.6(1.9)\cdot 10^{-5}$

\item On one occasion we observed the ordinary superhumps in quiescence
which seems to be common property of ER UMa stars.

\item No periodic light variations which can be connected with orbital
period of the binary were seen.

\item Striking stability of 19-day supercycle of RZ LMi and high activity
of the star may be caused by the presence of third body in the system.

\end{itemize}

\bigskip \noindent {\bf Acknowledgments.} ~We acknowledge generous
allocation of  the Warsaw Observatory 0.6-m telescope time.   Data from
AAVSO observers are also appreciated. We would like to thank Prof.
J\'ozef Smak for reading and commenting on the manuscript. KZ was
supported by the  Foundation for the Polish Science through grant
MISTRZ.


{\small

\begin{thebibliography}{}
   \bibitem{be00} Berry R., Burnell, J, 2000, The Handbook of Astronomical
           Imaging Processing, Willmann-Bell, Inc., Richmond, VA, USA.
   \bibitem{ga99} Gao W., Li Z., Wu X., Zhang Z., Li Y., 1999, ApJL, 527, L55
   \bibitem{gr82} Green R.F., Ferguson D.H., Liebert J., Schmidt M., 1982, PASP, 94, 560
   \bibitem{gr86} Green R.F., Schmidt M., Liebert J., 1986, ApJS, 61, 305
   \bibitem{he01} Hellier C., 2001, PASP, 113, 469
   \bibitem{he95} Henden A.A., Honeycutt R.K., 1995, PASP, 107, 324
   \bibitem{is01} Ishioka R., Kato T., Uemura M., Iwamatsu H.,
           Matsumoto K., Martin B.E., Billings G.W., Novak R., 2001,
           PASJ, 53, L51
   \bibitem{ka96} Kato T., Nogami D., Baba H., 1996, PASJ, 48, L93
   \bibitem{ka01} Kato T., 2001, PASJ, 53, L17
   \bibitem{ka03} Kato T., Nogami D., Masuda S., 2003, PASJ, 55, L7
   \bibitem{ls} Lipovetskij V., Stepanjan J., 1981, Astrophysics, 17, 573
   \bibitem{mi95} Misselt K.A., Shafter A.W., 1995, AJ, 109, 1757
   \bibitem{no95} Nogami D., Kato T., Masuda S., Hirata R.,
           Matsumoto K., Tanabe K., Yokoo T., 1995, 47, 897
   \bibitem{ol03} Olech A., Schwarzenberg-Czerny A., P. K\c{e}dzierski,
           K. Z{\l}oczewski, K. Mularczyk, M. Wi\'sniewski, 2003a, Acta
           Astron., 53, 175
   \bibitem{ol04} Olech A., Z{\l}oczewski K., Mularczyk K., K\c{e}dzierski P.,
           Wi\'sniewski M., Stachowski G., 2004, Acta Astron., 54, 57
   \bibitem{ol06} Olech A., Mularczyk K., K\c{e}dzierski P., Z{\l}oczewski K., 
           Wi\'sniewski M., Szaruga, K., 2006, Astron. Astrophys., 2006, 933
   \bibitem{ol07} Olech A., Rutkowski A., Schwarzenberg-Czerny A., 2007, 
           Acta Astron., 57, 331
   \bibitem{os95} Osaki Y., 1995, PASJ, 47, L11
   \bibitem{os96} Osaki Y., 1996, PASP, 108, 39
   \bibitem{pa95} Patterson J., Jablonski F., Koen C., O'Donoghue D.,
           Skillman D.R., 1995, PASP, 107, 1183
   \bibitem{pa98} Patterson J., 1998, PASP, 110, 1132
   \bibitem{pa01} Patterson J., 2001, PASP, 113, 736
   \bibitem{re02} Retter, A., Chou Y., Bedding T.R., Naylor T., 2002,
           MNRAS, 330, L37
   \bibitem{ro95} Robertson J.W., Honeycutt R.K., Turner G.W., 1995,
           PASP, 107, 443
   \bibitem{ru07} Rutkowski A., Olech A., Mularczyk K., Boyd D., Koff R., 
           Wi\'sniewski M., 2007, Acta Astron., 57, 267
   \bibitem{} Schreiber, M.R, Lasota, J.-P. 2007, arXiv:0706.3888
   \bibitem{sc96} Schwarzenberg-Czerny A., 1996, ApJ Letters,
           460, L107
   \bibitem{sp93} Skillman D.R., Patterson J., 1993, ApJ, 417, 298
   \bibitem{} Smak, J. 2000, {\it New Astronomy}, 44, 171
   \bibitem{st87} Stetson P.B., 1987, PASP, 99, 191
   \bibitem{sz92} Szkody P., Howell S.B., 1992, ApJS, 78, 537
   \bibitem{th97} Thorstensen J.R., Taylor C.J., Becker C.M., Remillard R.A., 
           1997, PASP, 109, 477
   \bibitem{th02} Thorstensen J.R., Patterson J., Kemp J., Vennes S.,
           2002, PASP, 114, 1108
   \bibitem{up92}  Udalski A., Pych W., 1992, Acta Astron.,
           42, 285
   \bibitem{war95} Warner B., 1995, {\it Cataclysmic Variable Stars}, 
           Cambridge University Press
   \bibitem{zh06} Zhao Y., Li Z., Wu X., Peng Q., Zhang Z., Li Z., 2006, PASJ, 58, 367
\end{thebibliography}
}
\end{document}